\documentclass[runningheads]{llncs}
\usepackage{amsmath, amssymb}
\usepackage{graphicx}
\usepackage{hyperref}
\usepackage{tikz}
\usetikzlibrary{positioning, shapes.geometric, arrows}
\usepackage{booktabs}

\begin{document}
	
	\title{Attention-Based Multimodal Survival Prediction with Cross-Modal Bilinear Fusion}
	\titlerunning{Multi-Modal Survival Prediction}
	
	\author{Hassan Keshvarikhojasteh, Josien P.W. Pluim, Mitko Veta}
	\authorrunning{Keshvarikhojasteh et al.}
	
	\institute{
		Department of Biomedical Engineering, Eindhoven University of Technology, Eindhoven, The Netherlands\\
		\email{h.keshvarikhojasteh@tue.nl}
	}
	
	\maketitle
	
	\begin{abstract}
		We propose a novel multimodal deep learning framework for patient-level survival prediction, which integrates whole-slide histology features, RNA-seq expression profiles, and clinical variables. Our architecture combines an ABMIL module~\cite{ilse2018attention} for slide-level representation with feedforward encoders for RNA and clinical data. These embeddings are then integrated through low-rank bilinear cross-modal fusion~\cite{liu2018efficient} to model conditional interactions across modalities while controlling parameter growth. The model outputs continuous risk scores that are subsequently mapped to survival times using a nonparametric calibration procedure based on the Kaplan–Meier estimator~\cite{kaplan1958nonparametric}. By decomposing multimodal reasoning into independent pairwise interactions, the proposed fusion design promotes structural interpretability and parameter efficiency compared with full tensor and hierarchical fusion strategies. Experiments on the CHIMERA challenge dataset demonstrate improved predictive performance over concatenation-based baselines and competitive generalization on hidden evaluation cohorts. These results indicate that the proposed framework is a promising approach for multimodal survival prediction in HR-NMIBC. The implementation is publicly available at \url{https://github.com/hassancpu/ChimeraChallenge2025_Task_3}.
	\end{abstract}

	\textbf{Keywords:} Multimodal survival prediction, Attention-Based Multiple Instance Learning (ABMIL), Cross-modal bilinear fusion, Kaplan--Meier estimation.

	\newpage
	\section{Introduction}
	Survival prediction from multimodal biomedical data is a key challenge in computational pathology and precision medicine. Histopathology whole-slide images (WSIs) capture tissue-level context, RNA-seq provides molecular signatures, and clinical features encode global patient characteristics. Combining these heterogeneous sources can improve robustness and predictive performance in survival outcome prediction~\cite{cheerla2019deep}. 
	
	However, the main challenge is the cross-modal integration of heterogeneous biomedical information, allowing the model to exploit complementary information between modalities. Simple concatenation of modality embeddings may overlook nonlinear dependencies. In contrast, bilinear fusion mechanisms can explicitly model pairwise multiplicative interactions, learning how specific gene expression patterns modulate the prognostic relevance of histological or clinical features~\cite{liu2018efficient}. This property is particularly critical for cancer prognostication, where interactions between morphology, transcriptomics, and clinical indicators often drive disease behavior~\cite{cheerla2019deep}.
	
	Attention-Based Multiple Instance Learning (ABMIL)~\cite{ilse2018attention} provides an interpretable mechanism to identify discriminative regions within histology slides by assigning adaptive weights to patch-level features. Compared to conventional pooling methods, the attention mechanism captures spatial heterogeneity and highlights diagnostically relevant tissue patterns, making it particularly suitable for survival tasks where prognosis often depends on localized morphological features.
	
	Several studies have explored multimodal survival prediction integrating imaging and molecular data. Cheerla and Gevaert (2019) proposed a deep multimodal architecture combining histopathology, clinical data, and transcriptomics, demonstrating the improved prognostic value of joint embeddings~\cite{cheerla2019deep}. More recent works~\cite{chen2022pan,chen2020pathomic} have expanded these approaches using cross-modal attention and bilinear pooling, achieving better interpretability and performance. Baltrušaitis et al.\ (2019) provide a comprehensive survey and taxonomy of multimodal machine learning, highlighting the importance of modality alignment and joint representation learning for capturing dependencies between heterogeneous data types~\cite{baltrusaitis2019multimodal}. Despite these advances, existing frameworks often rely on complex multi-stage training or extensive pretraining, limiting applicability in real-world scenarios.
	
	Our proposed method differs from prior multimodal survival models in three primary aspects: (1) it uses an ABMIL backbone for histology modeling, promoting interpretability and scalability; (2) it employs parameter-efficient low-rank bilinear fusion to explicitly capture interactions between modalities; and (3) it introduces a straightforward risk-to-time calibration mechanism based on empirical Kaplan–Meier estimates to obtain time-to-event predictions from model-derived risk scores.
	
\begin{figure}[!t]
		\centering
		\resizebox{\textwidth}{!}{
			\includegraphics{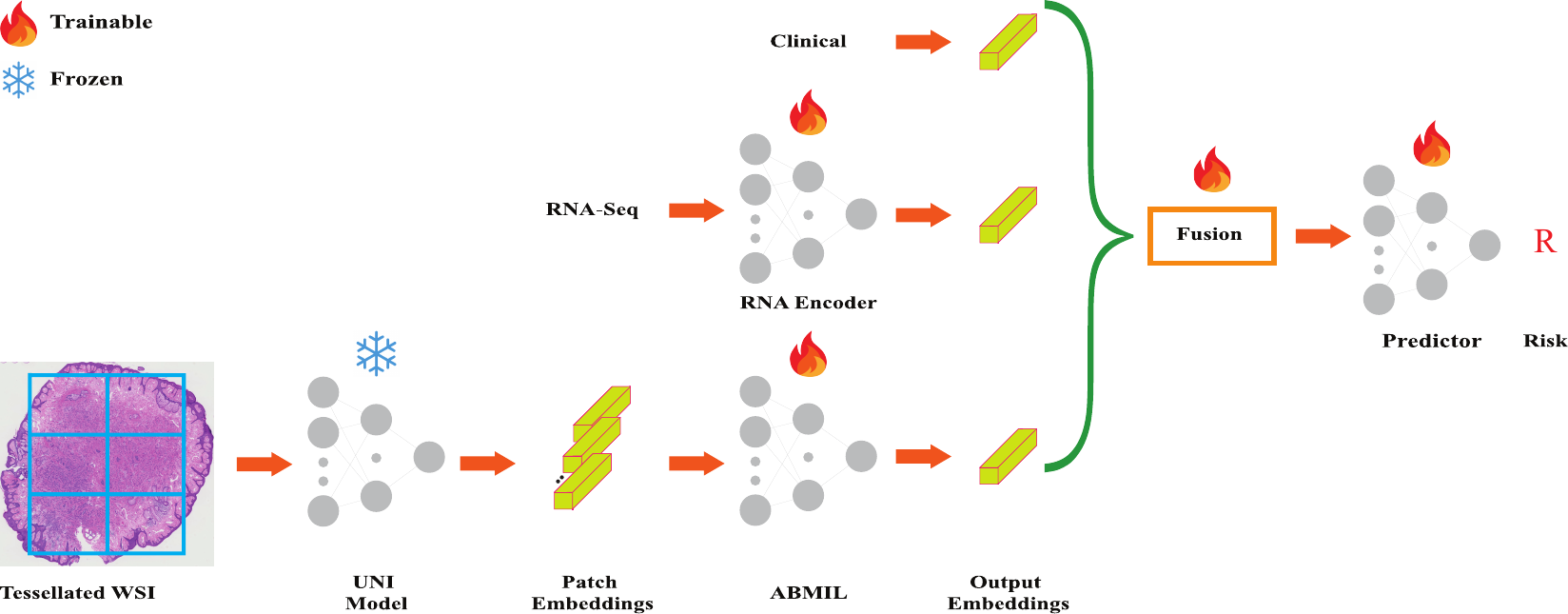}}
		\caption{Framework of the proposed method. Patch features are first extracted using the UNI model, and the slide representation is obtained using Attention-Based Multiple Instance Learning (ABMIL). The RNA-seq profile is compressed via the RNA encoder. The slide representation, compressed RNA embedding, and clinical data are then input to the fusion module. Finally, the risk score is predicted using a fully connected layer (predictor). For schematic clarity, all embeddings are depicted with similar box sizes, although their actual dimensionalities differ (1024-D for histology, 128-D for RNA, and 27-D for clinical features)}
		\label{fig:1}
	\end{figure}

	We designed the proposed framework with three guiding principles: interpretability, modularity, and efficiency. The histology encoder leverages ABMIL to isolate morphologically salient tissue regions. The RNA-seq and clinical branches distill high-dimensional information into compact embeddings through  feedforward encoders. The core novelty lies in our cross-modal fusion strategy, which systematically applies low-rank bilinear pooling across all modality pairs (histology$\times$RNA, histology$\times$clinical, RNA$\times$clinical) and ablates its interaction with residual connections. This comprehensive fusion analysis balances expressive power and computational cost while providing clear insights into cross-modal dependencies. The resulting joint embeddings support fine-grained survival risk estimation calibrated through nonparametric survival curve fitting. Figure~\ref{fig:1} illustrates the overall pipeline.
	
	In the following sections, we detail the architecture and training procedure, describe how it was applied to the non-muscle-invasive bladder cancer (NMIBC) dataset from the Combining HIstology, Medical imaging and molEcular data for medical pRognosis and diAgnosis (CHIMERA) Challenge, and report comparative results and discussion.
	
	\section{Materials \& Methods}
	
	\subsection{The CHIMERA Challenge}
	The CHIMERA Challenge is a MICCAI 2025 challenge that aims to advance precision oncology through the development and benchmarking of multimodal AI models integrating radiology, histopathology, molecular profiles, and clinical data. In contrast to earlier MICCAI challenges that typically focused on single-modality image segmentation or classification, CHIMERA explicitly targets the integration of heterogeneous data sources to improve cancer prognosis. The challenge comprises three tasks centered on prostate cancer and non-muscle-invasive bladder cancer (NMIBC), each formulated as a supervised prediction problem with clinically relevant endpoints~\cite{chimera_challenge,miccai_registered_challenges}.

	The third task (Task 3) concerns recurrence prediction in high-risk non-muscle-invasive bladder cancer (HR-NMIBC), requiring predicted survival times rather than raw risk scores. This task integrates histopathology WSIs, transcriptomics, and clinical covariates to model time-to-recurrence. By framing prognosis as a multimodal learning problem and relying on routinely collected clinical data, the challenge provides a realistic setting for developing multimodal survival models in line with precision oncology workflows~\cite{chimera_challenge}.

	\subsection{Problem Definition}
	The specific goal of Task 3 is to model time-to-recurrence for HR-NMIBC patients using multimodal data. Performance is quantified using the censored concordance index (C-index), measuring how well predicted risk scores preserve the observed temporal ordering of recurrence.
	
	\subsection{Dataset}
	The full CHIMERA Task 3 dataset consisted of 368 multimodal patient cases, which the organizers partitioned into a public training set and hidden validation and test sets for challenge evaluation. We used the public training portion, containing 176 multimodal patient cases. Each case includes haematoxylin and eosin (H\&E)-stained WSIs at 0.25 µm/pixel resolution, binary tissue masks, DESeq2-normalized~\cite{love2014moderated} RNA-seq profiles, and 27 structured clinical variables. Two cohorts were defined: Cohort A (3A; 126 cases) and Cohort B (3B; 50 cases), acquired using different experimental protocols. No batch effect correction was applied to retain real-world cohort heterogeneity, mimicking clinical validation scenarios. The ground-truth outcomes for each patient consisted of time-to-recurrence and event indicators, reflecting whether a recurrence occurred within the follow-up period. 
	
	\subsection{Histology Processing}
	Non-overlapping image patches of size 224 × 224 pixels are extracted at 10× magnification. Then, each patch is encoded into a 1024-dimensional feature vector using the pretrained UNI foundation model~\cite{chen2024towards}, selected for its state-of-the-art performance in computational pathology and compatibility with platform resource constraints. The ABMIL module with gated attention~\cite{ilse2018attention} aggregates these patch representations into a single slide-level embedding, enabling spatially aware weighting of discriminative tissue regions.
	
	\subsection{RNA Encoder}
	RNA-seq profiles, consisting of 19,360 features, are encoded using a feedforward neural network. The encoder begins with a linear transformation from the input space to a 2,048-dimensional hidden layer, followed by a ReLU activation that introduces nonlinearity and allows the model to capture complex interactions between genes. A dropout layer (0.25) is applied after each hidden layer to prevent overfitting. The second hidden layer reduces the dimensionality further to 512 features, again followed by ReLU and dropout. Finally, a linear projection maps the intermediate representation into a 128-dimensional embedding, which is passed through a ReLU activation to retain only positive, informative signals. Through this sequence of transformations, the encoder compresses the high-dimensional RNA-seq data into a compact, low-dimensional, and information-rich molecular representation suitable for downstream multimodal fusion. 
	
	\subsection{Clinical Features}
	Clinical variables (27 features) were processed to ensure comparability across patients and suitability for modeling. Numeric features, such as age and number of prior instillations, were standardized to have zero mean and unit variance. Categorical variables, including sex, smoking status, tumor stage, grade, and therapeutic indicators (e.g., Bacillus Calmette-Guérin (BCG) response group), were one-hot encoded to allow the model to treat each category independently. All preprocessing parameters were estimated using only the training cohort to prevent data leakage. 
	
	\subsection{Fusion Strategies}
	Fusion constitutes the central component of the proposed framework, as the prognostic signal in oncology often reflects conditional relationships between the underlying morphology, molecular alterations, and clinical context that each modality measures, rather than independent contributions of isolated unimodal predictors.

\begin{figure}[!t]
		\centering
		\includegraphics[width=0.7\linewidth]{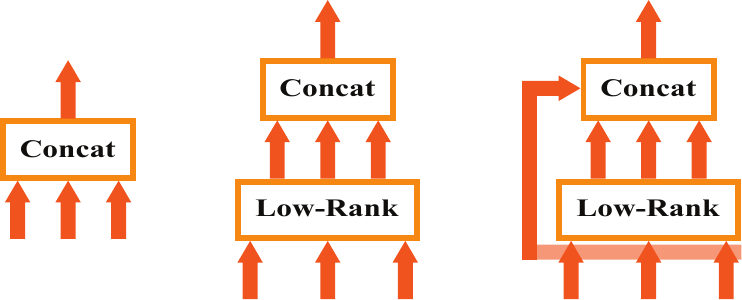}
		\caption{Different fusion modules. From left to right: naive concatenation; pair-wise fusion between modalities using low-rank bilinear followed by concatenation of the computed outputs; and concatenation with a residual connection.}
		\label{fig:2}
	\end{figure}

	We therefore investigate three increasingly expressive integration mechanisms while explicitly controlling model complexity to prevent overfitting in small patient cohorts, illustrated in Figure~\ref{fig:2}. We denote these models as ABMIL\_Surv, ABMIL\_Surv\_PG, and ABMIL\_Surv\_PG\_Res, where \emph{ABMIL} refers to the attention-based multiple instance learning histology encoder, \emph{Surv} indicates the survival prediction head, \emph{PG} denotes the use of pairwise (P) low-rank bilinear interactions constructed from global (G) modality embeddings, and \emph{Res} denotes the addition of a residual connection from raw modality embeddings.
	
	\paragraph{Concatenation baseline (ABMIL\_Surv)}
	The baseline model performs a straightforward concatenation of modality embeddings (histology, RNA, and clinical), which are then passed directly to a fully connected predictor. No normalization, scaling, or dimensionality alignment is applied prior to concatenation. This configuration assumes conditional independence between modalities and captures only additive contributions. While computationally simple, this approach cannot explicitly model interactions, such as a gene expression pattern altering the prognostic relevance of a morphological structure, highlighting the need for cross-modal fusion strategies.
	
	\paragraph{Full tensor and multimodal low-rank fusion (related work)}
	Prior multimodal fusion methods frequently rely on tensor-product interactions. Low-rank Multimodal Fusion (LMF)~\cite{liu2018efficient} approximates the full outer-product tensor across all modalities simultaneously. For three modalities $h$, $r$, and $c$, LMF learns a factorized representation of:
	
	\begin{equation}
		h \otimes r \otimes c
	\end{equation}
	
	capturing higher-order joint dependencies, where $\otimes$ denotes outer-product interaction. Although expressive, this formulation implicitly assumes sufficient data to learn stable multi-way correlations and often requires strong regularization.
	
	HFBSurv~\cite{li2022hfbsurv} instead applies bilinear pooling in a staged manner, progressively combining modalities in a hierarchical structure. While computationally more efficient than full tensor fusion, the hierarchical design introduces ordering dependencies between modalities and mixes modality-specific and interaction signals within intermediate representations.
	
	\paragraph{Proposed pairwise low-rank bilinear fusion (ABMIL\_Surv\_PG)}
	In contrast, we propose modeling cross-modal dependencies explicitly as \emph{pairwise interactions} rather than a single joint tensor interaction. We construct three independent low-rank bilinear modules:
	
	\begin{equation}
		(h \times r), \quad (h \times c), \quad (r \times c)
	\end{equation}
	
	Each module learns multiplicative relationships using a factorized bilinear mapping:
	
	\begin{equation}
		f(x,y) = W\left((U x) \odot (V y)\right)
	\end{equation}
	
	where $U$ and $V$ project modalities into a shared latent space and $\odot$ denotes element-wise interaction. The resulting interaction are concatenated before the survival predictor. 
	
	This design differs fundamentally from both LMF and hierarchical fusion approaches. While LMF approximates a single three-way correlation tensor across all modalities, our formulation decomposes multimodal reasoning into independent pairwise conditional interactions. This reduces parameter growth and relaxes the assumption that stable higher-order correlations can be learned from limited patient cohorts. Unlike hierarchical bilinear fusion (HFBSurv), which sequentially combines modalities and introduces ordering dependencies, our approach models all modality pairs in parallel. This order-independent design disentangles modality-specific and interaction-specific signals, improving interpretability and reducing interference between independent prognostic effects. A structured comparison of these strategies is summarized in 
	Table~\ref{tab:comparison}.
	
	\begin{table}[!t]
		\centering
		\caption{Comparison of multimodal fusion strategies highlighting interaction type, scalability, data requirements, and interpretability.}
		\label{tab:comparison}
		\resizebox{\textwidth}{!}{
			\begin{tabular}{lcccc}
				\toprule
				\textbf{Method} & \textbf{Interaction type} & \textbf{Scalability} & \textbf{Data requirement} & \textbf{Interpretability} \\
				\midrule
				Concatenation & additive & high & low & low \\
				LMF & full multi-way tensor & medium & high & low \\
				HFBSurv & hierarchical bilinear & medium & medium & medium \\
				\textbf{Ours} & pairwise conditional & \textbf{high} & \textbf{low} & \textbf{high} \\
				\bottomrule
		\end{tabular}}
	\end{table}
	
	\paragraph{Residual integration (ABMIL\_Surv\_PG\_Res)}
	To evaluate whether unimodal prognostic signals remain beneficial after interaction modeling, we introduce a residual pathway concatenating raw modality embeddings with the pairwise interaction features. This configuration allows the model to combine independent and conditional prognostic effects. 
	
	Our fusion strategy is motivated by clinical cohort characteristics, in which patient-level datasets are small relative to multimodal feature dimensionality and interpretability benefits from disentangling modality relationships. Accordingly, pairwise low-rank bilinear fusion provides a compromise between additive models and full tensor fusion, achieving expressive cross-modal reasoning.
	
	\subsection{Risk-to-Time Mapping}
	After training, the model outputs continuous risk scores for each patient, which indicate relative likelihood of recurrence but do not correspond directly to survival times. To translate these abstract scores into time-to-event predictions required by the challenge evaluation protocol, we apply a nonparametric mapping based on the Kaplan–Meier (KM) estimator~\cite{kaplan1958nonparametric}. First, patients in the training cohort are ranked according to their predicted risk, with higher-risk patients corresponding to earlier expected events. A KM curve is then fit using the observed times-to-event and censoring information, providing a mapping from survival probabilities to actual survival times.
	
	For a new patient, their predicted risk is compared to the distribution of training risks to determine an associated survival probability. The closest match on the KM curve is selected to produce a predicted survival time. This procedure preserves the relative ordering of patients while yielding clinically interpretable time-to-event predictions. Predicted survival times are clipped to non-negative values to maintain realistic estimates.

	\section{Experiments}
	
	\subsection{Implementation Details}
	Training was performed using the Cox proportional hazards objective~\cite{cox1972regression} and optimized with the Adam optimizer~\cite{kingma2014adam} (learning rate $1\times 10^{-3}$, weight decay $4\times 10^{-1}$, batch size 2, maximum 200 epochs). Dropout ($p=0.25$) and early stopping (30-epoch patience) were used for regularization. We conducted model development exclusively on the 176 public training cases. For this purpose, the dataset was randomly split at the patient level into training and internal validation sets using an 85:15 ratio, ensuring that all modalities from a given patient were assigned to the same split. Cohort A (3A) and Cohort B (3B) patients were preserved in their original proportions within both splits as much as possible, so that the internal validation set remained comparable in terms of cohort composition. To ensure consistent feature representation across slides, we adopted a fixed global gene ordering: a gene index was first constructed from a reference slide and stored, and all RNA expression vectors were then built by iterating over this fixed gene list and assigning the corresponding expression value (or zero if absent) for each gene. The hidden validation (external validation) and final test sets were defined and held back by the challenge organizers. Ground-truth labels for these hidden subsets were not available to participants; only aggregate performance metrics (C-index) were returned via the submission platform. Consequently, all hyperparameter tuning and model selection were based solely on the internal validation subset of the public training data, while the reported external validation and final test C-indices reflect independent evaluation on organizer-defined hidden cohorts. Model selection was performed by monitoring the validation performance and selecting the epoch achieving the maximum C-index. All experiments were implemented in PyTorch and executed on a single NVIDIA A30 GPU.

	\subsection{Results}
	
	\begin{table}[!t]
		\centering
		\caption{Censored concordance index (C-index) across training, internal validation, external validation, and final test sets. ABMIL\_Surv\_PG was selected for external/test evaluation due to superior internal validation performance and challenge submission limits. Where available, 95\% confidence intervals (CI) are reported for the internal validation C-index.}
		\label{tab:results_all_models}
		\resizebox{\textwidth}{!}{
			\begin{tabular}{lcccc}
				\toprule
				\textbf{Model} & \textbf{Training} & \textbf{Internal Val.} & \textbf{External Val.} & \textbf{Final Test} \\
				\midrule
				ABMIL\_Surv & \textbf{0.69} & 0.63 (0.43, 0.81) & -- & -- \\
				\textbf{ABMIL\_Surv\_PG} & 0.67 & \textbf{0.91 (0.82, 0.98)} & \textbf{0.66} & \textbf{0.68} \\
				ABMIL\_Surv\_PG\_Res & 0.53 & 0.89 (0.81, 0.96) & -- & -- \\
				\bottomrule
		\end{tabular}}
	\end{table}
	
	Table~\ref{tab:results_all_models} reports the C-index across training, internal validation, external validation, and final test sets for all three fusion strategies. The bilinear fusion model ABMIL\_Surv\_PG achieved the highest internal validation C-index of 0.91, substantially outperforming the concatenation baseline ABMIL\_Surv (0.63). Notably, ABMIL\_Surv\_PG demonstrated superior generalization with only a modest training C-index of 0.67, highlighting the effectiveness of low-rank bilinear modeling for cross-modal interactions in heterogeneous biomedical data.
	
	The residual variant ABMIL\_Surv\_PG\_Res showed reduced training performance (C-index = 0.53) but maintained strong validation results (0.89). We hypothesize that this is due to signal dilution in the residual pathway, where raw modality embeddings introduce noise that interferes with the refined bilinear interaction terms. This behavior, together with the small sample size, suggests that the additional flexibility of the residual branch may promote overfitting. Altogether, these observations highlight the importance of explicitly modeling cross-modal dependencies in small datasets while avoiding unnecessarily complex fusion mechanisms.
	
	ABMIL\_Surv\_PG was selected for external evaluation due to challenge submission limits permitting only one model. This model maintained competitive generalization with external validation (0.66) and final test set performance (0.68), achieving a C-index slightly below the challenge winner's score of 0.68~\cite{chimera_challenge}.
	
	These findings demonstrate four key insights: (1) explicit pairwise bilinear fusion substantially improves survival prediction over simple concatenation; (2) low-rank bilinear modeling provides an effective balance between expressiveness and generalization; (3) residual connections may dilute rather than enhance cross-modal signal quality in this setting; and (4) the proposed fusion strategy achieves near state-of-the-art performance relative to the challenge winner.
	
	\begin{figure}[!t]
		\centering
		\resizebox{\textwidth}{!}{
			\includegraphics{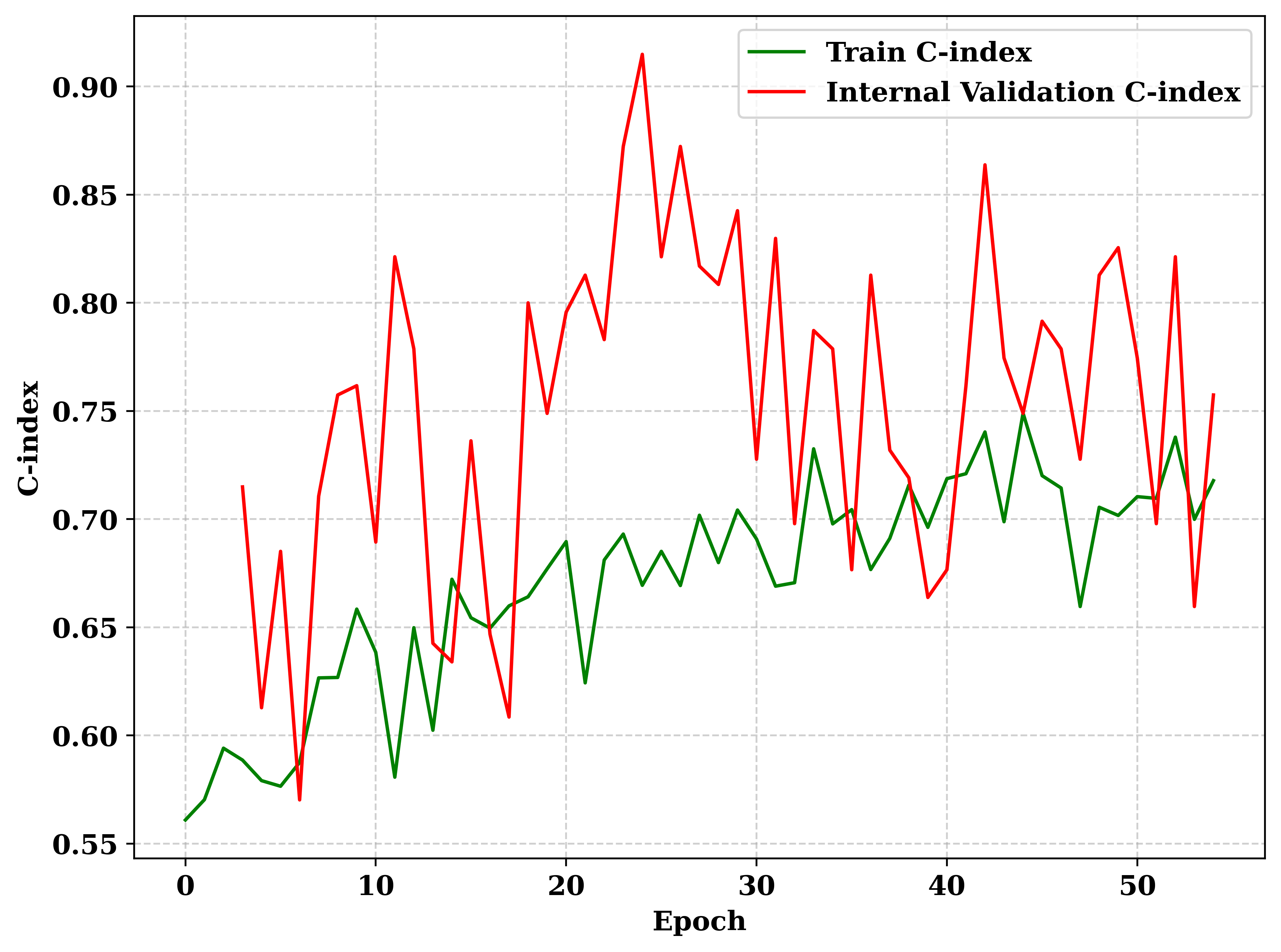}}
		\caption{Training and internal validation censored concordance index (C-index) of ABMIL\_Surv\_PG\ across epochs.}
		\label{fig:3}
	\end{figure}
	
	\begin{figure}[!t]
		\centering
		\resizebox{\textwidth}{!}{
			\includegraphics{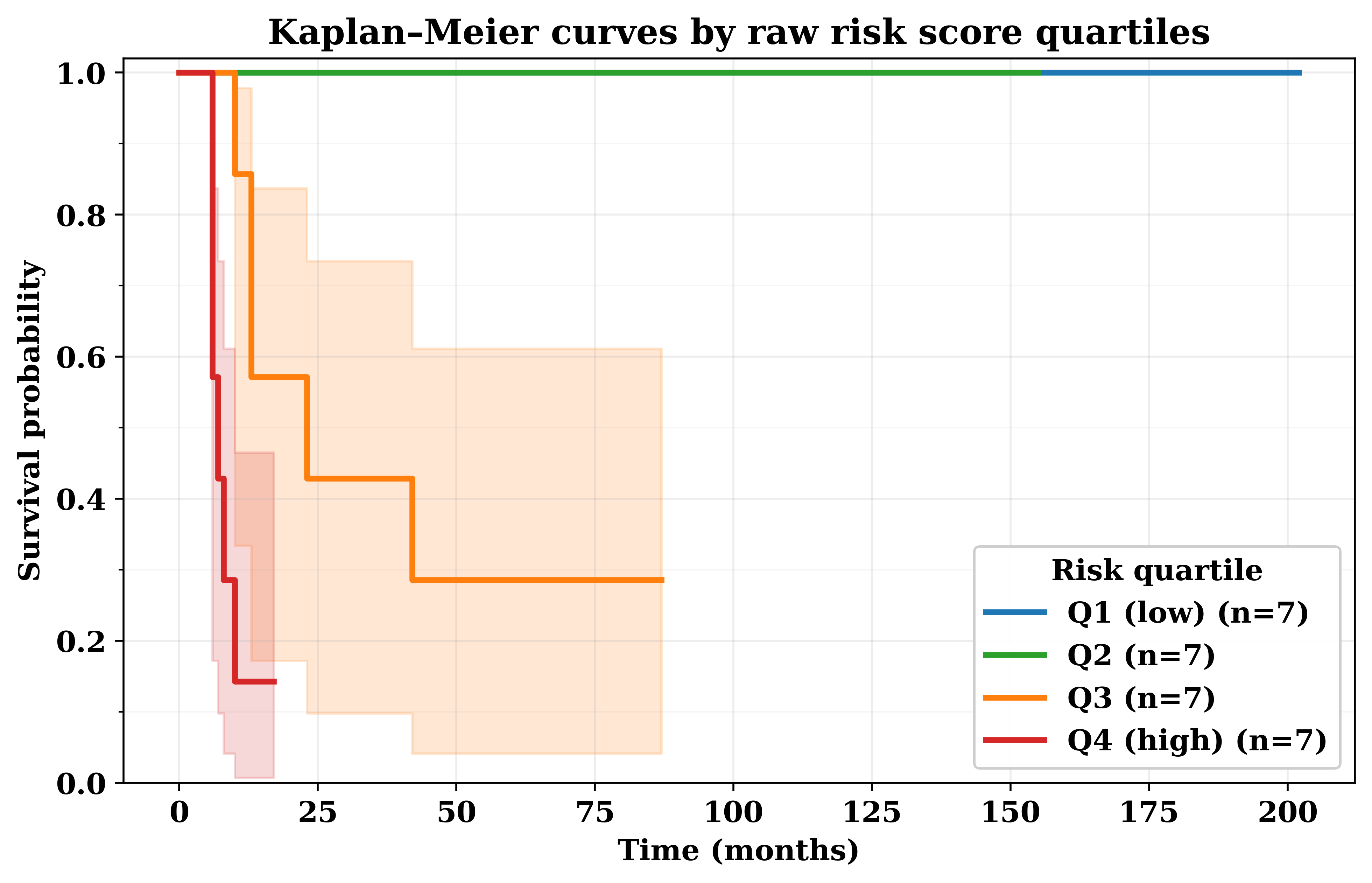}}
		\caption{Kaplan–Meier survival curves on the internal validation set, stratified by quartiles of the predicted raw risk score (Q1: lowest risk, Q4: highest risk).}
		
		\label{fig:4}
	\end{figure}
	
	The training C-index dynamics of ABMIL\_Surv\_PG are illustrated in Figure~\ref{fig:3}. The plot reveals overfitting after approximately 23 epochs, as evidenced by the divergence between training and validation curves, which reflects the limited dataset size of 176 patients. Figure~\ref{fig:4} demonstrates highly effective risk stratification based on the continuous risk scores: patients in the highest risk quartile (Q4) experience rapid recurrence with median survival of approximately 13 months (50\% survival probability), dropping to about 18\% event-free survival by 48 months. By sharp contrast, patients in the lowest risk quartile (Q1) maintain 100\% event-free survival throughout the entire 205-month observation period. This dramatic separation across risk-score quartiles quantitatively validates the model's ability to transform raw risk estimates into prognostically meaningful groups, distinguishing patients at imminent risk of recurrence from those with excellent long-term prognosis. Although the small number of patients per group (n=7) leads to wide confidence intervals and statistically unstable estimates, the monotonic trend across quartiles indicates that higher predicted risk is associated with systematically poorer observed survival.
	
	\section{Discussion and Conclusion}
	The proposed multimodal survival prediction framework demonstrates that combining histology, RNA-seq, and clinical data through attention-based aggregation and bilinear cross-modal fusion substantially improves prognostic accuracy over simpler fusion approaches. The ABMIL encoder effectively localized prognostically relevant tissue regions, offering interpretable insights into morphological correlates of recurrence risk. The bilinear fusion modules captured critical higher-order cross-modal dependencies between histology, RNA-seq, and clinical features, outperforming naïve concatenation under equal model capacity.

	Interpretability and modularity are key advantages of this framework. Attention maps derived from ABMIL can highlight tissue regions associated with poor prognosis, while the fusion architecture allows disentangling cross-modality contributions. Furthermore, the model requires minimal pretraining or parameter tuning, which simplifies deployment in research settings and facilitates future translational studies in biopsy-driven prognostic workflows.
	
	Nevertheless, model generalizability remains a limitation. The lack of batch correction across cohorts introduces domain shifts that could hinder performance in external datasets. Future work will extend this approach by introducing domain-adaptive fusion mechanisms and uncertainty modeling for out-of-distribution detection. Additionally, extension to longitudinal and multi-tumor datasets will be explored, alongside interpretability analysis linking attention weights and gene expression signatures to biological pathways.
	
	Overall, the proposed ABMIL-based multimodal fusion model has been demonstrated as a scalable and high-performing framework for survival prediction from heterogeneous biomedical data, with competitive performance in international benchmarking challenges on HR-NMIBC. While additional validation on independent cohorts is required before clinical implementation, the presented results provide evidence that pairwise bilinear fusion is an effective and practically useful design choice for multimodal survival modeling in this setting.
	
	\section*{Acknowledgments}
	This work was supported by the IMI BigPicture project (IMI945358). We thank our colleages at the Medical Image Analysis Group (IMAG/e) for their encouragement during the submission peroid.
	
	\bibliographystyle{splncs04}
	\bibliography{reference}
	
\end{document}